\documentstyle[12pt]{article}

\def\simlt{\stackrel{<}{{}_\sim}}

\def\Eq{\begin{equation}}
\def\End{\end{equation}}

\def\Eqa{\begin{eqnarray}}
\def\Enda{\end{eqnarray}}

\def\Endl#1{\label{#1} \End}

\def\ord#1{{\cal O}(#1)}

\begin{document}
\begin{titlepage}
\title{MODEL INDEPENDENT UPPER BOUND ON THE LIGHTEST HIGGS
BOSON MASS IN SUPERSYMMETRIC STANDARD MODELS\thanks{Based on talk
given at the XXVIIIth Rencontres de Moriond: {\it Electroweak
Interactions and Unified Theories}, 13--20 March, 1993}}

\vspace{1cm}

\author{{\bf Mariano Quir\'os}\thanks{Work partly supported by CICYT, Spain,
under contract AEN90-0139.} \\
Instituto de Estructura de la Materia, CSIC \\
Serrano 123, 28006-Madrid, Spain}

\vspace{2cm}
\date{}
\maketitle
\begin{abstract}
One of the main features of the Minimal Supersymmetric Standard Model (MSSM)
is the existence of an absolute tree-level upper bound $m_h$ on the mass
of the $CP=+1$ lightest Higgs boson, equal to $m_Z$, that could affect
detectability at future colliders. The above bound is spoiled by
{\bf radiative corrections} and by an {\bf enlarged Higgs sector},
as {\em e.g.} a
gauge singlet. Radiative corrections in the MSSM can push the upper
bound up to $115\ GeV$ for $m_t \simlt 150\ GeV$. The presence of an enlarged
Higgs sector changes the previous upper bound to one depending on
the electroweak scale, $\tan \beta$ and the gauge and Yukawa couplings
of the theory. When radiative corrections are included, the allowed region
in the $(m_h,m_t)$ plane depends on the scale $\Lambda$ below which the
theory remains perturbative. In particular, for models
with  arbitrary Higgs sectors
and couplings saturating the scale $\Lambda=10^{16}\ GeV$ we find
$m_h \simlt 155\ GeV$ and $m_t \simlt 190\ GeV$.
\end{abstract}
\thispagestyle{empty}

\vskip-17.cm
\noindent
\phantom{bla}
\hfill{{\bf hep-ph/9305243}} \\
\phantom{bla}
\hfill{{\bf IEM-FT-72/93}}
\end{titlepage}
\section{Minimal Supersymmetric Standard Model}

Supersymmetric models have well constrained Higgs sectors \cite{SUSY}
which can provide crucial tests of them. In particular, the MSSM
contains two Higgs doublets $H_1$, $H_2$, with hypercharges
$Y=\pm 1/2$, coupled to quarks and leptons in the superpotential
\Eq
f_m=h_t  Q\cdot H_2 U^c + h_b  Q\cdot H_1 D^c+
h_{\tau} L\cdot H_1 E^c,
\Endl{fm}
and a physical Higgs sector containing two scalar ($h,H$), one
pseudoscalar ($A$) and two charged ($H^{\pm}$) states.
Its most constraining feature is
the existence of an absolute upper bound on the tree-level
mass of the $CP=+1$ lightest Higgs boson
$h$ given by (with $\tan\beta\equiv v_2/v_1$,
$v_i\equiv\langle H_i^o\rangle$),
\Eq
m_h\leq m_Z\mid \cos 2\beta\mid \ .
\Endl{mhmssm}

The bound (\ref{mhmssm}) would suggest that the light Higgs of the MSSM
should be discovered at LEP-200 \cite{LEP2}. So, what would happen if
the Higgs is {\em not} discovered with a mass below $m_Z$? There is
still a hope for supersymmetry since the bound (\ref{mhmssm}) is
spoiled by {\bf radiative corrections} and by the presence of an
{\bf enlarged Higgs sector}. The question to be answered in this talk is,
how large are those effects?

Radiative corrections in the MSSM
have been computed using different methods, with results
which are in very good agreement to each other:
{\it i)} Standard diagrammatic techniques \cite{SDT};
{\it ii)} One-loop effective potential \cite{EP}; and,
{\it iii)} Renormalization group approach (RGA) \cite{RGA,EQ1}.
The latter (RGA) method is {\it reliable} for a scale of supersymmetry
breaking $\Lambda_s^2 \gg m_W^2$. The technical reason being that in the
limit $\Lambda_s^2/m_W^2 \rightarrow \infty$ the Higgs
$h \rightarrow (\cos \beta H_1^{or}+\sin \beta H_2^{or})$ becomes the
Standard Model Higgs, with a mass $m_h= m_Z\mid~ \cos 2\beta~\mid$,
while the other states decouple, since
$m_H \sim m_A \sim m_{H^{\pm}} \sim \Lambda_s$.
The RGA method is also {\it universal} in the sense that it
is valid (can pick the leading radiative contribution from the top-stop
sector) for general supersymmetric standard models. We have
shown in Fig.~1 the upper bound on $m_h$ in the MSSM
for different values of $\Lambda_s$, from $1$ to $10\ TeV$,
using the RGA method of Ref. \cite{EQ1}, which includes two-loop corrections.
{}From Fig.~1 one can conclude that for $\Lambda_s \simlt 10\ TeV$ and
$m_t \simlt 150\ GeV$, $m_h \simlt 115\ GeV$.
%
%
\section{MSSM with singlets}
The case of the MSSM with a gauge singlet field $S$
coupled to the rest of the Higgs sector through the
cubic superpotential
\Eq
f_h=\lambda S\ H_1\cdot H_2+\frac{1}{3}\chi S^3
\Endl{fh}
was first studied in \cite{DREES,ELLIS} where the tree level bound
\Eq
m_h^2\leq \left(\cos^2 2\beta +\frac{2\lambda^2\cos^2
\theta_W}{g^2}\sin^2 2\beta\right)m_Z^2,
\Endl{mhsing}
was found. Notice that (\ref{mhsing}) does not depend either on the
supersymmetric masses or on the supersymmetry breaking terms.
It depends on $\lambda$, whose maximum
value ($\lambda^{{\rm max}}$) is obtained assuming the
theory remains perturbative below the high scale $\Lambda \sim
\Lambda_{{\rm GUT}}$ \cite{EQ2}-\cite{EL}. We have included $h_t$ and
$h_b$ in (\ref{fm}) with the boundary condition $m_b(10\ GeV)=5\ GeV$.
For a fixed value of $\tan\beta$ ({\it i.e.} $h_b$), $\lambda$ and $h_t$
are related to each other through the renormalization group equations
(RGE), as shown in Fig.~2a.

Radiative corrections are introduced using different methods
(the RGA \cite{EQ3,EQ4,EKW}, and the one-loop effective
potential \cite{COMELLI,E}) with results consistent within less than
5\% for all the considered $m_t$-range. The final upper bound on
$m_h$ is shown in Fig.~2b from where one can see
that the absolute upper bound
is $m_h \simlt 145\ GeV$, though its precise
value depends on $m_t$ and $\tan\beta$.
\section{General supersymmetric standard models}

The previous analysis has been generalized in Refs.\cite{EQ3,EQ4}
to supersymmetric standard models with arbitrary Higgs sectors.
We have introduced, on top of the Higgs doublets $H_1$ and $H_2$, which
take vacuum expectation values and are coupled to the quarks and leptons
through (\ref{fm}):
{\it i)} An arbitrary number of extra pairs $H_1^{(j)}$,
$H_2^{(j)}$, $j=1,...,d$, decoupled from quarks and leptons in
order to avoid dangerous flavor changing neutral currents;
{\it ii)} Gauge singlets $S^{(\sigma)}$, $\sigma=1,...,n_s$;
{\it iii)} $SU(2)$ triplets $\Sigma^{(a)}$, $a=1,...,t_o$, with $Y=0$;
{\it iv)} $SU(2)$ triplets $\Psi^{(i)}_1$, $\Psi^{(i)}_2$,
$i=1,...,t_1$, with $Y=\pm 1$.
Other Higgs representations will only contribute to the gauge
$\beta$-functions and will not provide renormalizable couplings to
$H_1 \cdot H_2$, $H_1 \cdot H_1$ and $H_2 \cdot H_2$. They will
result in lower values of the upper bounds and need not be considered.

Assuming the most general renormalizable superpotential generalizing
(\ref{fh})
\Eq
f_h=\vec{\lambda}_1 \vec{S}H_1 \cdot H_2+
\vec{\lambda}_2 H_1 \cdot \vec{\Sigma} H_2+
\vec{\chi}_1 \cdot \vec{\Psi}_1 H_1 +
\vec{\chi}_2 \cdot \vec{\Psi}_2 H_2 +
\ord{\Sigma^3,\Sigma \Psi_1 \Psi_2, S^3}
\Endl{fhgen}
one can obtain the tree-level bound
\Eq
m_h^2/v^2\leq\frac{1}{2}(g^2+g'^2)\cos^2 2\beta +
(\vec{\lambda}^2_{1}
+\frac{1}{2}\vec{\lambda}^2_{2})\sin^2 2
\beta+ \vec{\chi}^2_1\cos^4\beta+ \vec{\chi}^2_2\sin^4\beta,
\Endl{mhgen}
where $v^2\equiv v_1^2+v_2^2$ and $g,g'$ are the $SU(2)\times U(1)_Y$
gauge couplings. In particular, the bound (\ref{mhsing}) is recovered when
$\vec{\lambda}_{2}=\vec{\chi}_{1}=\vec{\chi}_{2}=0$, and the bound
(\ref{mhmssm}) in the MSSM when also $\vec{\lambda}_{1}=0$.
The bound (\ref{mhgen}) is
independent of the soft-supersymmetry breaking
parameters, or any supersymmetric mass terms. It
is only controlled by $v$ and dimensionless parameters
($g,g',\tan\beta$, and Yukawa couplings). Since the former is
fixed by the electroweak scale, the latter will determine the bound
(\ref{mhgen}). The upper bound then
comes from the requirement that the supersymmetric theory remains
perturbative below $\Lambda$. In particular the bound
is maximized when some of the involved couplings
(generically $\lambda$) {\it saturate} the
high-scale $\Lambda$, {\it i.e.} when $\lambda^2(\Lambda)/4\pi\sim 1$.
To guarantee this condition we need to
solve the RGE of all gauge and Yukawa couplings of the theory.

Given a model $m\equiv (n_s,d,t_0,t_1)$, the gauge couplings saturate a
particular scale $\Lambda_m$. Conversely, given a scale $\Lambda$ there is
a set of models whose gauge couplings saturate $\Lambda$. Out of them we
pick the model which provides the maximum value of the upper bound.
There is in this way a one-to-one correspondence between models and scales,
and we can obtain the upper bound $m_h(\Lambda)$ for all models which are
perturbative below the given scale $\Lambda$.
For instance the scale $\Lambda=10^{16}\ GeV$ is saturated by the
model $n_s>0$, $d=5$ (see also Ref. \cite{KKW}).
For simplicity, we have analyzed the case $t_0=t_1$,
where $\rho_{{\rm tree}}=1$ by tuning vacuum expectation values.

Radiative corrections are introduced using the RGA \cite{EQ3,EQ4} with
$\Lambda_s=1\ TeV$. The upper bound corresponding to
$\Lambda=10^{16}\ GeV$ is shown in Fig.~3a. We can see from
Fig.~3a that
the absolute upper bound is $m_h \simlt 155\ GeV$ for all values of
$m_t$ and $\tan\beta$, though there is a strong dependence on those
parameters. The upper bound $m_h(\Lambda)$ as a function of $m_t$, and any
value of $\tan\beta$, is shown in Fig.~3b for different values of the
scale $\Lambda$.
\section{Conclusions}
We have obtained the most general upper bound for the mass
of the lightest Higgs boson in supersymmetric standard models which
remain perturbative below a scale $\Lambda$. Radiative corrections
are taken into account using the renormalization group approach.
For a given scale the obtained bounds are stronger than those in the
Standard Model and satisfy $m_h(\Lambda)\longrightarrow
(m_h)_{{\rm SM}}$ for $\Lambda \rightarrow G_F^{-1/2}$.
Our results are not taking into account (and could be affected by):
{\it i)} The presence of {\it extra colored fermions} ($t',b',\ldots$) that
could introduce new radiative corrections \cite{MOROI};
{\it ii)} The presence of an {\it extra gauge group factor}
($g_a,T_a$) that would result in $\Delta m_h^2=\ord{g_a^2}$.

\section*{Acknowledgements}
The content of this talk is based on work done in
collaboration with Jose Ramon Espinosa to whom
I want to express my deep gratitude. I also thank G. Kane and
F. Zwirner for useful discussions.

\newpage
\section*{Figure captions}
\begin{description}
\item[Fig.1]
Upper bounds on $m_h$ in the MSSM
\item[Fig.2]
$\lambda^{{\rm max}}$ and $m_h$ in the MSSM with a singlet
\item[Fig.3]
$m_h(10^{16} GeV)$ and $m_h(\Lambda)$ for general
supersymmetric standard models
\end{description}
\end{document}